\documentclass[twocolumn,prb,showpacs]{revtex4}

\usepackage{graphicx}
\usepackage{rotating}
\usepackage{amsmath}
\usepackage{amsfonts}
\usepackage{amssymb}
\usepackage{enumerate}
\usepackage{longtable}
\usepackage{dcolumn}% Align table columns on decimal point
\usepackage{bm}

\begin{document}
\newcommand{\sm}{${\rm SmOFeAs}$}
\newcommand{\smx}{${\rm SmO_{1-x}F_{x}FeAs}$}
\newcommand{\be}{\begin{equation}}
\newcommand{\ee}{\end{equation}}
\newcommand{\bn}{\begin{eqnarray}}
\newcommand{\en}{\end{eqnarray}}

\title{Mottness underpins the anomalous optical response of Iron
Pnictides}

\author{M. S. Laad,$^1$ L. Craco,$^{2,3}$ S. Leoni,$^2$ and H. Rosner$^2$}
\affiliation{$^1$Max-Planck-Institut f\"ur Physik Komplexer Systeme.
01187 Dresden, Germany \\
$^2$Max-Planck-Institut f\"ur Chemische Physik fester Stoffe,
%N\"othnitzer Strasse 40,
01187 Dresden, Germany \\
$^3$Leibniz-Institut f\"ur Festk\"orper- und Werkstoffforschung Dresden,
%Helmholtz Strasse 20,
01069 Dresden, Germany}

\date{\rm\today}

\begin{abstract}
The recent discovery of high-temperature superconductivity (HTSC) in doped 
Iron pnictides is the latest example of unanticipated behavior exhibited
by $d$- and $f$-band materials.  The symmetry of the SC gap, along with the 
mechanism of its emergence from the ``normal'' state is a central issue
in this context.  Here, motivated by a host of experimental signatures 
suggesting strong correlations in the Fe-pnictides, we undertake a detailed
study of their normal state. Focussing on symmetry-unbroken phases, we use
the correlated band structure method, LDA+DMFT, to study the one-particle 
responses of both ${\rm LaO_{1-x}FeAsF_{x}}$ and ${\rm SmO_{1-x}FeAsF_{x}}$ 
in detail. Basing ourselves on excellent quantitative agreement between 
LDA+DMFT and key experiments probing the one-particle responses, we extend 
our study, undertaking the first detailed study of their normal state 
electrodynamic response. In particular, we propose that near-total normal 
state incoherence, resulting from strong, {\it local} correlations in the 
Fe $d$-shell in Fe-pnictides, underpins the incoherent normal state 
transport found in these materials, and discuss the specific electronic 
mechanisms leading to such behavior. We also discuss the implications of 
our work for the multi-band nature of the SC by studying the pairing 
``glue'' function, which we find to be an overdamped, electronic continuum. 
Similarities and differences between cuprates and Fe-pnictides are also 
touched upon.  Our study supports the view that SC in Fe-pnictides arises 
from a bad metallic, incoherent ``normal'' state that is proximate to a 
Mott insulator.  
\end{abstract}


\maketitle

\section{Introduction}

Discovery of high-$T_{c}$ superconductivity (HTSC) in the Fe-based
pnictides~\cite{[1]} is the latest among a host of other, ill-understood
phenomena characteristic of doped $d$- and $f$-band compounds. HTSC in
Fe-pnictides emerges upon doping a bad metal with spin density wave (SDW) 
order at ${\bf q}=(\pi,0)$. Preliminary experiments indicate~\cite{[2],[3]}
unconventional SC. Extant normal state data indicate a ``bad metal''
with anomalously high resistivity $O$(m$\Omega$-cm) even at low
temperature.~\cite{[1]} These observations in Fe-pnictides are
reminiscent of {\it underdoped} cuprate SC. The small carrier density,
along with Uemura scaling from $\mu$-SR~\cite{[1]} similar to hole-doped
cuprates strongly suggests a SC closer to the Bose condensed, rather
than a BCS limit.  A brief general review~\cite{[dirk]} gives a chemist's 
overview on the subject.

Several theoretical works have addressed the issue of the ``degree of
correlated electronic'' behavior in Fe-pnictides.~\cite{Haule,si,shori}
This is an important issue, bearing as it does upon a characterization of
charge and spin fluctuations: are they itinerant,~\cite{kuroki,ere} or 
closer to localized?~\cite{Haule,si,bask} This itinerant-localized 
{\it duality} is a recurring theme in correlated systems in 
general,~\cite{imada} and in fact is at the root of early formulations 
of the Hubbard model itself.~\cite{hubbard}

In Fe-pnictides, HTSC results from the Fe $d$ band states hybridized with
As $p$ states: this leads to two hole, and two electron-like
pockets~\cite{[5]} in one-electron bandstructure calculations. Within
weak coupling HF-RPA studies of {\it effective}, two- and four-orbital
Hubbard models,~\cite{raghu,ere} this gives a ${\bf q}=(\pi,0)$ SDW order,
in seeming agreement with inelastic neutron scattering (INS)
results.~\cite{[7]} Observation of quasi-linear (in $T$) behavior in
the resistivity, pseudogap in optical reflectance,~\cite{[8]} and a spin
{\it gap} in NMR~\cite{[3]} in {\it doped} Fe-pnictides, among other 
observations, however, are benchmark features showing the relevance of 
strong, dynamical spin and charge correlations in the pnictides. In 
analogy with cuprates, this suggests that the Fe-pnictides might be 
closer to a Mott insulator (MI) than generally thought.~\cite{si} 
Actually, the undoped pnictides of the ${\rm A_{1-x}FeAsF_{x}}$ type 
with $A=La,Sm$, the so-called ``$111$'' pnictides, show an insulator-like 
resistivity {\it without} magnetic order for $T^{*}>137$~K,~\cite{[1]} 
dependent upon the specific Fe-pnictide considered. Onset of bad metallic 
behavior correlates with a structural (tetragonal-orthorhombic (T-O)) 
distortion at $T^{*}$, {\it below} which SDW order sets in.  This is 
different from the ${\rm AFe_{2}As_{2}}$ pnictides with $A=Ba,Sr$ (the 
so-called ``$122$'' pnictides), where the bad metallic resistivity is 
observed only above $T^{*}$. Nevertheless, as we will discuss below, 
optical measurements on the $122$ family also indicate a non-Fermi Liquid 
(nFL) metallic behavior at low $T$. So due care must be exercised when 
one attempts to classify Fe-pnictides into the ``weakly'' or ``strongly''
correlated category. The small carrier number seemingly generated upon 
the structural distortion in the $111$ pnictides accords with the observed 
high resistivity, lending further credence to such a view.

Optical conductivity is a time-tested probe for characterizing the charge
dynamics in solids.  Specifically, it measures how a particle-hole pair
excitation, created by an external photon field, propagates in the system,
uncovering the detailed nature of the excitation spectrum itself. In a
normal Fermi liquid (FL), low-energy scattering processes leave the identity
of an excited electron (hole) intact.  This fact, noticed first by Landau,
implies a long lifetime for excited {\it quasiparticles}: near the Fermi
surface, their lifetime, $\tau^{-1}(\omega,T)\simeq \omega^{2},T^{2}$,
greatly exceeds $\omega,T$, their energy. In optical response, this fact
manifests itself as the low-energy Drude part (after subtracting the phonon
contribution), corresponding to coherent propagation of particle-hole
excitations built from such quasiparticles. The Drude parametrization,
a lorentzian with half-width $\Gamma=\tau^{-1}(\omega,T)$

\be
\sigma(\omega)=\frac{ne^{2}}{m^{*}}\frac{1}{1+i\omega\tau(\omega)}
\ee
describes the low-energy optical response of normal metals. This allows
one to estimate the transport relexation rate and dynamical mass
from~\cite{millis}

\be
\frac{1}{\tau(\omega)}=
\frac{Ne^{2}}{m_{0}} Re \left[ \frac{1}{\sigma(\omega)} \right]
\ee
and
\be
m^{*}(\omega)=
\frac{Ne^{2}}{\omega} Im \left[ \frac{1}{\sigma(\omega)} \right] \;.
\ee
With $\sigma(\omega) \simeq \Gamma/(\omega^{2}+\Gamma^{2})$ at low-energy,
$m^{*}(\omega)=m_{0}$, a constant. As long as the FL survives, even in
$f$-band rare-earth metals with $e-e$ interactions much larger than the
(band) kinetic einergy, this observation holds. Observation of a non-Drude
optical conductivity in clean metals with low residual resistivity is
thus a diagnostic for non-FL charge dynamics, i.e, where
$\tau^{-1}(\omega)\simeq\omega^{2}, m^{*}=const$ at low energy no longer
hold.  One can, however, continue to use the Drude parametrization, at
the cost of having a complicated $\omega$-dependence of $\tau^{-1},m^{*}$
at low energy.  Such non-FL optical conductivity in the
symmetry-{\it unbroken} bad-metallic phases is characteristic of several
strongly correlated systems, from quasi-one dimensional Luttinger
liquids,~\cite{organ} high-$T_{c}$ cuprates up to optimal 
doping,~\cite{vandm} $f$-electron systems close to quantum phase 
transitions~\cite{geibel} and MnSi,~\cite{vandm1} among others. 
Additionally, strongly correlated $d$- and $f$-band FL metals routinely 
exhibit a non-Drude optical response above a low-$T$ scale, the so-called 
lattice {\it coherence} scale ($O(1-20)$~K), below which correlated FL 
behavior obtains.~\cite{imada} So the material diversity and range of 
distinct ground states exhibited by the above strongly suggests a common 
underlying origin of the anomalous charge dynamics.  The known importance 
of strong, short-ranged electronic correlations in $d$-and $f$-band systems 
then implies that Mott-Hubbard physics may underlie such generic, anomalous 
features.
   
On the theoretical side, observation of non-Drude, incoherent, or power 
law optical response forces one to discard the Landau FL theory, together 
with perturbation theory in interactions upon which it is based: an 
electron (hole) is no longer an elementary excitation of the system. The
one-fermion propagator exhibits a {\it branch cut} analytic structure,
leading to power-law fall-off in optics. This reflects the fact that the
action of the electronic current operator (within linear response
theory)~\cite{shastry} does {\it not} create well-defined elementary
excitations at low energy, leading to an incoherent response.

\section{Earlier Work}

Extant LDA+DMFT (local-density approximation plus dynamical mean-field
theory) works on Fe-pnictides give either a strongly renormalized
FL~\cite{Haule} or an orbital selective (OS), incoherent,
pseudogapped metal.~\cite{shori} Very good semiquantitative agreement
with key features seen in {\it both} photoemission (PES) and X-ray
absorption (XAS) for ${\rm SmO_{1-x}F_{x}FeAs}$,~\cite{pes,xas} as well 
as with the low-energy ($15$ meV) kink in PES is obtained using
LDA+DMFT.~\cite{SmFe} Focusing on the ``normal'' state of Fe-pnictides,
is LDA+DMFT adequate for describing their {\it correlated} electronic
structure, or are cluster extensions (cluster-DMFT) of DMFT needed? If
the observed SDW order has its origin in a Mott-Hubbard, as opposed to
a weak-coupling Slater-like SDW picture, one would expect that
incorporation of ``Mottness''~\cite{philips} is adequate, at least in
the symmetry unbroken phases ($T>T^{*}$ at $x=0$ and at all $T>T_{c}$, the
SC transition temperature, for doped cases) without SDW/SC order.  If 
two-particle spectra, e.g, the optical conductivity, could be described 
within the {\it same} picture, this would serve as strong evidence for 
relevance of large $D$ (DMFT) approaches in this regime.  Specifically, 
given that vertex corrections in the Bethe-Salpeter equations for the 
conductivity identically vanish in $D=\infty$,~\cite{khurana} a proper 
description of the optical response of ${\rm SmO_{1-x}F_{x}FeAs}$ within 
DMFT would imply negligible vertex corrections, justifying use of DMFT 
{\it a posteriori}.

An optical study on La- and Sm-oxypnictides has already been carried
out.~\cite{drech,vdm} While detailed spectral weight analysis remains to
be done, characteristic strong correlation features are already visible:
a small ``Drude'' peak, weak mid-infra red feature, and a slowly decreasing
contribution up to high energy, $O(2.0)$~eV in La-pnictides, and a
power-law like decay of the reflectance in Sm-pnictide, all testify
to this fact, and accord with their bad metallic resistivity. Onset of
SC in ${\rm SmO_{1-x}FeAsF_{x}}$ results in reflectivity changes over a
broad energy range,~\cite{vdm} a characteristic signature of underlying
``Mottness''.  Apart from these common features, there are quantitative
differences in results from different groups.~\cite{drech,vdm} These can
be traced back to the fact that while an effective medium approximation
is invoked for ${\rm SmO_{1-x}FeAsF_{x}}$,~\cite{vdm} no such analysis is
performed for ${\rm LaO_{1-x}FeAsF_{x}}$.~\cite{drech} Given the intrinsic
polycrystal nature of the samples used by both, as well as the differences
in analysis, this may not be surprising. Optical work on single-crystal
samples is thus highly desirable; this may be close at hand. Nevertheless,
with these caveats, these observations are strongly reminiscent of cuprates
up to optimal doping,~\cite{imada} and constrain theories to
understand SC as an instability of an incoherent, non-FL metallic state.
More recently, Yang {\it et al.}~\cite{timusk} have indeed performed
a detailed analysis of optics for single crystals of
${\rm Ba_{0.55}K_{0.45}Fe_{2}As_{2}}$. Slow (in energy, $\omega$) decay in
reflectivity and an anomalously large $\tau^{-1}(\omega)$ with sub-linear 
$\omega$ dependence, implying no FL quasiparticles, reinforce similar 
features found in earlier optical results for La- and Sm-oxypnictides. 
Further, an extraction of the $\alpha^{2}F(\Omega)$ bosonic 
``glue''~\cite{timusk} reveals strong coupling to the low-energy bosonic 
fluctuation modes (due to inter-orbital, coupled charge-spin density 
modes?). Observation of non-FL quasiparticle features in optics implies 
that these bosonic modes are themselves strongly overdamped. Strong
inelastic scattering from short-ranged, dynamical spin, charge and orbital
correlations can indeed lead to such behavior, as has been investigated
more extensively in the cuprate context. More studies are required to
check whether these features are generic for Fe-pnictides: in view of
incoherent features already seen in all investigated cases, we believe
that this will indeed turn out to be the case.
    
In what follows, we compute the {\it correlated} band structure and optical
conductivity of {\it both} La- and ${\rm SmO_{1-x}FeAsF_{x}}$, extending
previous work, where very good semiquantitative agreement with the
one-particle spectrum was found.~\cite{SmFe} We show how ``Mottness'' in
the Fe $d$-bands underpins the charge dynamics in Fe-pnictides. In
particular, we show how an excellent theory versus experiment comparison
for the one-particle spectral function (DOS) is obtained for doped
${\rm LaO_{1-x}FeAsF_{x}}$, and build upon this agreement to obtain very
good {\it quantitative} agreement with the {\it reflectivity} as well.  
Armed with this agreement, we analyze these theoretical results in detail
and predict specific {\it non-FL} features that should be visible
in future experimental work. Finally, we estimate the ``glue function'',
$\alpha^{2}F(\omega)$, and propose that it should be understood as an
electronic (multiparticle) continuum that can be interpreted as an
overdamped bosonic spectrum. We conclude with a brief qualitative
discussion of its implications for SC.
 
\section{Correlated band structure}

\begin{figure}
%\begin{center}
\includegraphics[width=\columnwidth]{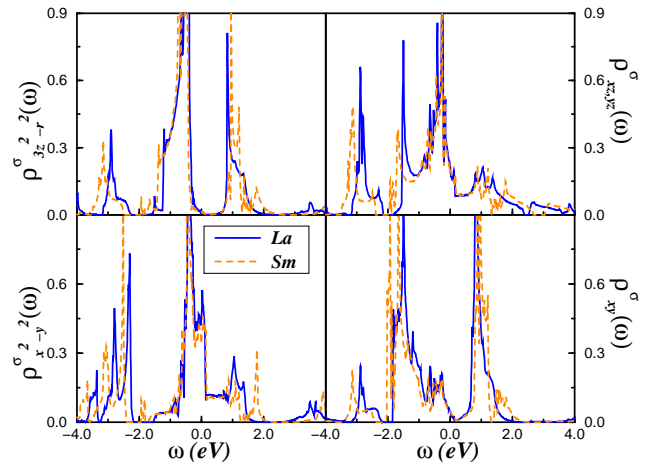}
%\includegraphics[width=3.1in]{F0-opt.eps}
%\end{center}
\caption{(Color online) Orbital-resolved LDA density-of-states (DOS) for 
the Fe $d$ orbitals in ${\rm LaOFeAs}$ (solid) and ${\rm SmOFeAs}$ (dashed), 
computed using the LMTO method.  Notice the overall similarity between the 
two DOSs. This shows that the electronic states generically relevant to 
Fe-pnictides are Fe $d$-band states.}
\label{fig1}
\end{figure}

The bare one-electron band structure of both ${\rm LaOFeAs}$ and
${\rm SmOFeAs}$ used in this work was computed using the linear 
muffin-tin orbital (LMTO) scheme~\cite{ok} in the atomic sphere 
approximation (ASA).~\cite{ASA} As evidenced in several studies,
the overall structures are remarkably similar in both cases, confirming
that the active electronic states involve the carriers in the FeAs layers.
Shown in Fig.~\ref{fig1}, the orbital-induced anisotropies in
the band structure are manifest: the $xy,3z^{2}-r^{2}$ bands are almost
gapped at the Fermi energy ($E_{F}$), while the $xz,yz,x^{2}-y^{2}$
bands have appreciable weight at $E_{F}$. The only essential difference 
between the LDA DOS for La- and Sm-pnictides is that in the latter case, 
due to larger chemical pressure (caused by the smaller size of Sm relative 
to La), the LDA band width is slightly ($O(0.5$~eV)) larger.  The 
one-electron part of the Hamiltonian for Fe-pnictides is then given by,

\be
H_{0}=\sum_{{\bf k},a,\sigma}(\epsilon_{{\bf k},a}-\epsilon_{a})
c_{{\bf k},a,\sigma}^{\dag}c_{{\bf k},a,\sigma}\;,
\ee
where $\epsilon_{{\bf k},a}$ label the five $d$ bands, and $\epsilon_{a}$
denote the band energies.  The inter-orbital splitting arises from the real
crystal field (of $S_{4}$ symmetry in Fe-pnictides), which lifts the
five-fold degeneracy of the atomic $d$-shell. This gives the two hole- and 
two electron-like pockets, as apparently observed by de Haas van Alphen
(dHvA) studies.~\cite{[hussey]}

However, a direct comparison between LDA results and the PES/XAS
experiments (which must be considered {\it together} when a comparison 
of the theoretical spectral function is attempted) shows substantial 
mismatch between theory and experiment. Related discrepancies are found 
in optical studies,~\cite{drech} where the actual plasma frequency, 
$\omega_{p}$, is $2-3$ times smaller than the LDA prediction. Neither 
can the high ($O$(m$\Omega$-cm)) resistivity be understood within an 
almost band-like (free electron) picture. Also, the dHvA 
study~\cite{[hussey]} reveals that the LDA bands need to be shifted by 
$0.2$~eV to get a proper fit with experiment. Further, the effective 
masses are enhanced by a factor of $2-3$ over their LDA values (this 
agrees with the renormalization of the plasma frequency found in optics). 
Taken together, these features strongly suggest sizable electronic 
correlations, which, moreover, are also exhibited by a host of other, 
known, correlated metals.~\cite{imada}

Theoretically, while LDA (LDA+U) generically accounts for {\it ground state}
properties of weakly (strongly) correlated systems, their inability to
describe excited states (and hence the charge and spin dynamics) in
correlated systems is well-documented.~\cite{kot-rev} Marrying LDA with
DMFT opens the way toward resolving this shortcoming of traditional band
structure approaches, and LDA+DMFT has proven successful in describing 
physical properties of various correlated materials in terms of their {\it
correlated} electronic structures.~\cite{kot-rev}

The discussion above clearly shows that incorporation of strong, multi-orbital
electronic correlations is a requirement for a proper description of Fe 
pnictides. The interaction part is given by,

\bn
\nonumber
H_{int} &=& U\sum_{i,a}n_{ia\uparrow}n_{ia\downarrow}
+ U'\sum_{i,a\ne b,\sigma,\sigma'}n_{ia\sigma}n_{ib\sigma'} \\
&-& J_{H}\sum_{i,a,b}{\bf S}_{ia} \cdot {\bf S}_{ib} \;,
\en  
where $n_{a\sigma}=c_{a\sigma}^{\dag}c_{a\sigma}$ and ${\bf S}_{a}$ are the
fermion number and spin density operators for an electron in orbital $a$.
We take $U \simeq U'+2J_{H}$, as is commonly known for TMO.~\cite{kot-rev}
The relevance of strong correlations in Fe-pnictides has been recognized by
several authors.~\cite{bask,si,Haule} While these works explicate the 
important role of multi-orbital (MO) correlations (in particular, the 
sensitivity to $J_{H}$),~\cite{Haule} other works~\cite{shori} conclude 
that correlations are weak. This is a highly relevant, and open, issue 
in the field of Fe-pnictides and their SC: are they weakly correlated, 
itinerant metals, with a conventional, BCS like instability to SC, or are 
they strongly correlated metals, giving way to SC via a non-BCS-like 
instability? A detailed comparison with extant experimental results 
should go a long way toward resolving this important issue.

Here, guided by good success obtained in a theory-experiment comparison of
one-particle spectra in our previous study,~\cite{SmFe} we use LDA+DMFT to 
compute the detailed optical response in the ``normal'' state of La- and 
Sm-based Fe-pnictides.  We solve $H=H_{0}+H_{int}$ within multi-orbital 
(MO) DMFT. In this study, MO-IPT is empolyed as the ``impurity solver'' 
to solve the impurity model of DMFT.  Though not numerically ``exact'', 
it has been shown to be quantitatively accurate for a wide range of 
correlated $d$-band materials.~\cite{[mllc],jap-paper}
If only the FeAs layer states are relevant in Fe-pnictides,~\cite{[5]}
we expect our work to provide a generic picture of charge dynamics in
Fe-pnictides.  We find excellent quantitative agreement with {\it both},
the one-particle spectral data (PES/XAS) as well as two-particle data 
(reflectance) for the La-based Fe-pnictide. Based on this, we argue 
that Fe-pnictides should be viewed as strongly correlated, MO systems, 
with incoherent low-energy behavior and describe the optical response 
of {\it both} La- and Sm-based Fe-pnictides in detail. Implications of 
our work for the high-$T_{c}$ SC in Fe-pnictides are touched upon, and 
intriguing similarities (and differences) with cuprates are highlighted.

\begin{figure}
%\begin{center}
\includegraphics[width=\columnwidth]{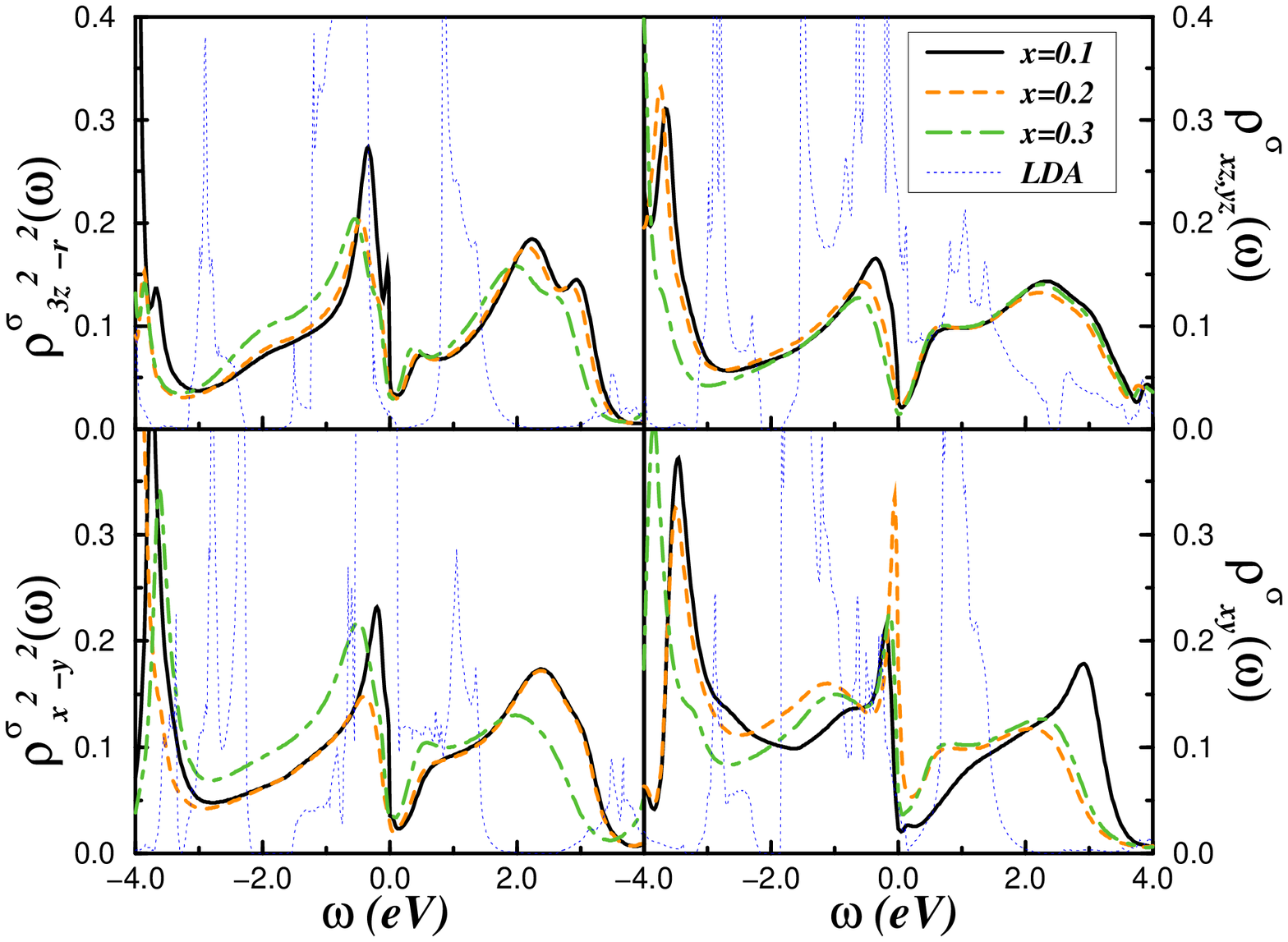}
\includegraphics[width=\columnwidth]{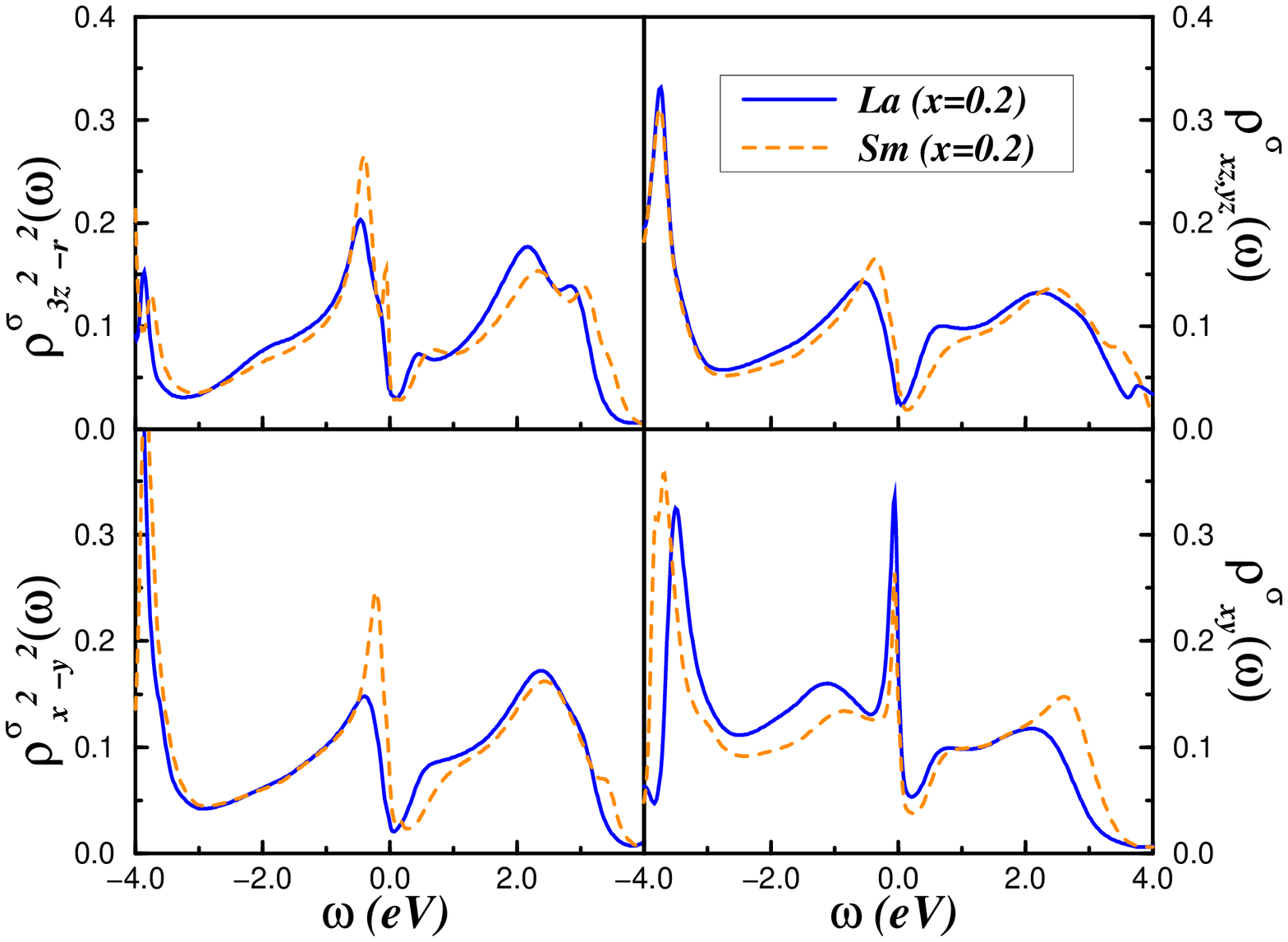}
%\includegraphics[width=3.1in]{F1b-opt.eps}
%\end{center}
\caption{(Color online) Top panel: Orbital-resolved LDA+DMFT (solid, dashed 
and dot-dashed) and LDA (dotted) DOS for electron-doped ${\rm LaOFeAs}$ for 
three doping values.  The parameters are $U=4.0$~eV, $U'=2.6$~eV and 
$J_{H}=0.7$~eV. The drastic modification of the LDA spectra to almost 
totally incoherent character by large-scale dynamical spectral weight 
transfer is clearly visible. Bottom panel: Comparison between LDA+DMFT 
spectra for La- and Sm-based Fe-pnictides.  
%Apart from their overall similarity, the more pronounced pseudogap features 
%for the La-pnictide, caused by a narrower LDA band width (compared 
%to Sm-pnictide), are clearly resolved.
}
\label{fig2}
\end{figure}

Starting with the five Fe $d$-orbitals, we use MO-DMFT to extract the 
correlated spectral functions for the five $d$-orbitals.  We choose values 
of $U=4.0$~eV, $J_{H}=0.7$~eV and $U'\simeq (U-2J_{H})=2.6$~eV, as employed 
in our earlier work.~\cite{SmFe} These are shown in Fig.~\ref{fig2} for 
La-based Fe-pnictide, for three values of electron doping, so that 
$n_{total}=\sum_{a}n_{a}=(6+x)$, with $x=0.1, 0.2, 0.3$, along with the 
respective LDA DOS. Electronic correlations are seen to lead to dramatic 
and interesting modifications of the LDA spectra:

$(i)$ the spectra describe an incoherent, non-FL metal for {\it each} of
the $d$-bands, with orbital dependent low-energy pseudogap features.  
Correspondingly, the imaginary parts of the self-energies (not shown)
show deviations from  the $-\omega^{2}$ form at small $\omega$, being
consistent instead with a sub-linear $\omega$-dependence, along with a
{\it finite} value at $E_{F}(=0)$, as also seen in our earlier
work.~\cite{SmFe}

$(ii)$ In striking contrast to the LDA band structures, the LDA+DMFT 
band structure shows that multi-orbital electronic correlations
``self-organize'' the spectral functions.  While the $xy$-orbital DOS 
shows the maximum itinerance, and has a shape distinct from the others, 
the much more localized $xz,yz,x^{2}-y^{2},3z^{2}-r^{2}$ orbital-DOS are 
seen to closely resemble each other with regard to their lineshapes.  
Dramatic spectral weight transfer (SWT) over large energy scales $O(5.0)$~eV 
is also apparent in the results.  In our MO-DMFT calculation, strong, 
incoherent inter-orbital charge transfer leads to dramatic spectral weight 
redistribution between the different $d$-orbital DOS.  This is a 
characteristic also exhibited by other, correlated, MO 
systems,~\cite{[bavs3]} and points to the relevance of MO correlations 
in the Fe-pnictides.

$(iii)$ There is no OS metallic phase for our choice of parameters, and
{\it all} $d$-orbital DOS cross $E_{F}$. The DMFT results are sensitive
to changes in $U,U'$ for fixed $J_{H}$, and our results describe a metal 
very close to the OS-metallic one, which occurs for 
$U \ge 5.0$~eV.~\cite{shori}

Very similar features have been reported by us for Sm pnictides in an 
earlier study.~\cite{SmFe} This is shown in the lower panel of 
Fig.~\ref{fig2}. Given that the LDA bands for Sm-based Fe-pnictide are 
slightly wider than for La-based Fe-pnictides, we expect localization-like 
features to be more enhanced in La-based Fe-pnictides within a Mott-Hubbard 
picture, as is indeed seen in the comparison. These results imply that 
strong, MO correlations may have a {\it generic} consequence of 
self-organizing the correlated spectra, an observation {\it not} readily 
seen in the LDA DOS.  This may have important consequences, and, for 
example, could be of aid when one seeks to construct effective 
{\it correlation} based models.

\begin{figure}
%\begin{center}
\includegraphics[width=\columnwidth]{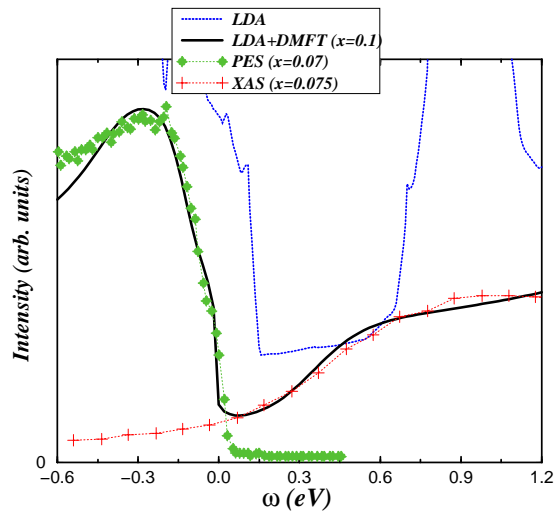}
%\includegraphics[width=3.1in]{F2-opt.eps}
%\end{center}
\caption{(Color online) Comparison between the LDA+DMFT result for 
${\rm LaO_{0.9}FeAsF_{0.1}}$ (solid line) and angle-integrated 
photoemission (PES, diamonds)~\cite{lpes} and X-ray absorption 
(XAS, vertical crosses)~\cite{lxas} for ${\rm LaO_{0.93}FeAsF_{0.07}}$. 
As in ${\rm SmOFeAs}$,~\cite{SmFe} very good quantitative agreement is 
clearly seen. In particular, the low-energy kink at $-15.0$~meV in PES 
is accurately resolved in the DMFT spectrum. Also notice the substantial 
improvement obtained by LDA+DMFT over the LDA result (dotted line).}
\label{fig3}
\end{figure}

In Fig.~\ref{fig3}, we compare our LDA+DMFT results with extant
photoemission (PES) and X-ray absorption (XAS) data for
${\rm LaO_{0.93}FeAsF_{0.07}}$.~\cite{lpes,lxas} Since only the five 
$d$-bands have been included in the LDA+DMFT, we restrict
ourselves to the energy window (which, however, is rather wide)
$-0.6\le\omega\le 1.2$~eV around $E_{F}$ (this is the region where only
the five $d$-bands dominate in the LDA).  Clearly, excellent quantitative
agreement with {\it both} PES and XAS results is obtained.  In particular,
the low-energy pseudogap is faithfully reproduced, as is the detailed form
of the lineshapes.  Taken together with our earlier results on
${\rm SmO_{1-x}FeAsF_{x}}$,~\cite{SmFe} this strongly suggests that the 
${\rm FeAs}$ states with sizable $d$-band electronic correlations are 
a universal feature of Fe-pnictides.

In particular, a noteworthy fact is that for {\it both} Fe-pnictides, a
low energy kink at approximately $15.0-25.0$~meV, along with a pseudogap,
and strongly asymmetric incoherent features at higher energies (at $-0.28$~eV
in PES and at $0.6$~eV in XAS) are clearly resolved.~\cite{lpes,lxas}  
Remarkably, our LDA+DMFT calculation reproduces {\it all} these features
in excellent agreement with both PES and XAS results.  The low-energy kink
is interpreted as arising from low-energy,  collective inter-orbital
fluctuations, as discussed earlier~\cite{SmFe} in detail.  

Additionally, two more interesting features are also visible from 
Fig~\ref{fig3}.  First, a good ``fit'' between the experimental and LDA 
values of the Fermi energy is achieved by shifting the LDA spectrum {\it
downwards} by $0.15$~eV.  The need for such a shift of LDA DOS in this 
context has already been noticed earlier~\cite{lpes} and attributed 
to correlation effects.  Hitherto, their quantification has not been 
undertaken. Here, this already arises selfconsistently in the MO-DMFT 
from the MO-Hartree shift mentioned above, bringing the LDA+DMFT value 
of $E_{F}$ in very good agreement with experiment. Second, as we will 
estimate below in optical analysis, the effective plasma frequency is 
reduced by a factor of $2-3$ over its LDA value. This translates into 
an {\it average} effective enhancement of the band (LDA) mass as 
$2-3m_{0}$, where $m_{0}$ is the bare LDA mass. {\it Both} these 
observations are completely consistent with the dHvA results, where 
very similar estimates for the band shift as well as the effective mass 
were extracted.~\cite{[hussey]} In addition, the latter is also consistent 
with the $2-3$-fold enhancement in the $\gamma$ co-efficient of the low-$T$ 
specific heat in ${\rm LaO_{1-x}FePF_{x}}$.~\cite{[spec-heat]} Taken 
together, these results constitute a consistent, {\it quantitative} 
rationalization of basic one-particle responses in both (La- and Sm-based) 
pnictides.          

\section{Optical conductivity using LDA+DMFT}

We now study the optical conductivity of both, doped ${\rm LaOFeAs}$ and
${\rm SmOFeAs}$ using the LDA+DMFT propagators for all $d$-orbitals.  In
$D=\infty$, the computation of the optical conductivity simplifies
considerably. This is because the irreducible vertex functions entering
the Bethe-Salpeter equation for the evaluation of the current-current
correlation function vanish {\it exactly} in this limit.~\cite{khurana}  
Thus, the optical response is directly evaluated as convolution of the 
DMFT propagators.  For MO-systems, the general expression for the real 
part of the optical conductivity is given by

\bn
\nonumber
\sigma_{ab}^{\prime} (\omega) &=&
\frac{2\pi e^{2}\hbar}{V} \sum_{\bf k}\int d\omega'
\frac{f(\omega')-f(\omega+\omega')}{\omega} \\ &\times&
Tr[{\bf A}_{\bf k}(\omega'+\omega)v_{{\bf k},a}{\bf A}_{\bf k}(\omega')
v_{{\bf k},b}] \;,
\en
with $a,b$ labelling the various orbitals used in the DMFT calculation.
${\bf A}_{\bf k}(\omega)$ is the one-particle spectral function, a
matrix in the orbital sector, 
% $V$ is volume of the unit cell per formula unit,
and $v_{{\bf k},a}=\langle{\bf k}|P_{a}|{\bf k}\rangle$
is the fermion velocity in orbital $a$.  The corresponding matrix element of
the momentum, $P_{a}$, weights the different transitions, and is determined
by the band structure.  Estimation of the $v_{{\bf k},a}$ for Fe-pnictides
is especially difficult, where, in addition to the $d$-states (which can be
written in localized, Wannier-like basis sets), one also has the much more
delocalized As $p$-states to contend with.  Hence, we simplify our analysis
by replacing this matrix element by a constant,
$v_{{\bf k},a}=\langle{\bf k}|P_{a}|{\bf k}\rangle = v_{a}$. Further,
we restrict ourselves to {\it intraband} transitions.~\cite{pava} Both 
these approximations will turn out to be justified later. With these
simplifications, the optical conductivity is written as

\bn
\nonumber
\sigma_{ab}^{\prime} (\omega) &=& \delta_{a,b} v_{a}^{2}
\frac{2\pi e^{2}\hbar}{V}\sum_{\bf k}\int d\omega'
\frac{f(\omega')-f(\omega+\omega')}{\omega} \\ &\times&
A_{{\bf k},a}(\omega'+\omega)A_{{\bf k},a}(\omega')\;,
\label{optc}
\en
where 

\be
A_{a}({\bf k},\omega)=-\frac{1}{\pi} Im \left[ 
\frac{1}{\omega-\epsilon_{{\bf k},a}-\Sigma_{a}(\omega)} \right]
\ee
is the fully renormalized one-particle spectral function for
orbital $a$, and $\Sigma_{a}(\omega)$ is the corresponding one-particle
self-energy. The total reflectivity, $R(\omega)=\sum_{a}R_{a}(\omega)$,
can be computed using

\be
R_{a}(\omega)= \left| \frac{\sqrt{\epsilon_{a}(\omega)} -1}
{\sqrt{\epsilon_{a}(\omega)} +1} \right|^2 \;,
\ee
with 

\be
\epsilon_{a}(\omega)=1+\frac{4\pi i\sigma_{a}(\omega)}{\omega}
\ee
being the complex dielectric constant.

Using the Kramers-Kr\"onig (KK) relations, a detailed analysis of the
reflectivity, $R(\omega)$, and the optical conductivity can be readily 
carried out for the normal state.  In addition, an extended Drude
analysis of the results shines light on the nature of the strong
renormalizations caused by MO electronic correlations.  In particular,
estimation of the frequency-dependent carrier lifetime, $\tau(\omega)$,
and effective mass, $m^{*}(\omega)$, for each orbital state yields
microscopic information on the mechanism of incoherent metal formation.
Finally, the electronic ``glue'' responsible for pairing is estimated
therefrom as,~\cite{millis}

\be
{\cal F}_{a}(\omega)\equiv\alpha_{a}^{2}F_{a}(\omega)=
\frac{1}{2\pi}\frac{d^{2}}{d\omega^{2}}
\left( \frac{\omega}{\tau_{a}(\omega)} \right)  
\ee
whence microscopic information concerning the detailed structure of the
pairing interaction is obtained.

\section{Results for electrodynamic response}

We now describe our results. In doing so, we adopt the following strategy:

$(i)$ we use our LDA+DMFT results for doped ${\rm LaOFeAs}$, which give a
very good description of the one-particle spectral features,
Fig.~\ref{fig2}, as discussed above. Using the DMFT propagators, we compute 
the {\it intraband} optical conductivity of ${\rm LaO_{1-x}FeAsF_{x}}$.  
Excellent quantitative agreement with extant reflectivity data as
measured~\cite{drech} is obtained, and, building upon  this agreement,
we describe the optical response (i.e, optical conductivity, dielectric
constants, plasma frequency, ac penetration depth) in detail.

$(ii)$ Given the observation that the electronic structure of the
Fe-pnictides is determined by the electronic states in the FeAs layers,
we use our DMFT results for {\it both} La- and Sm-based Fe-pnictides to
compute the normal state electrodynamics in the correlated metal.  We
provide the {\it first} theoretical estimates of the anisotropic carrier
lifetimes and effective masses as a function of frequency.  These results
show, in accord with the incoherent metal classification of single layer
Fe-pnictides, that their normal state cannot be described within FL
theory.  

\begin{figure}
%\begin{center}
\includegraphics[width=\columnwidth]{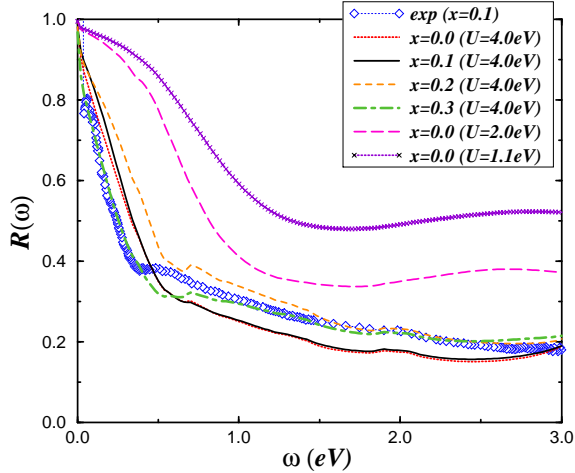}
%\includegraphics[width=3.1in]{F3-opt.eps}
%\end{center}
\caption{(Color online) Comparison between the experimental reflectivity 
for ${\rm LaO_{0.9}FeAsF_{0.1}}$~\cite{drech} and the LDA+DMFT results
for $n_{total}=(6+x)$. Very good quantitative agreement over the whole 
energy range, including the kink in the relectvity around $0.6$~eV is 
clearly seen for $U=4.0$~eV and $x=0.1$. Also, progressive {\it disagreement}
between theory and experiment with decreasing $U$ is clear: for 
$U=1.1,2.0$~eV, substantial disagreement over the whole energy range is 
strong evidence for the strongly correlated character of the metallic 
state in Fe-pnictides.}
\label{fig4}
\end{figure}

In Fig.~\ref{fig4}, we show the theory-experiment comparison for the
{\it reflectivity} of ${\rm LaO_{0.9}FeAsF_{0.1}}$. The experimental result
was taken from earlier work.~\cite{drech} Quite remarkably, excellent
semiquantitative agreement is clearly visible for $U=4.0$~eV,$U'=2.6$~eV, 
and $J_{H}=0.7$~eV in LDA+DMFT. To highlight the importance of strong
electronic correlations, we have also plotted theoretical results for 
smaller, unrealistic values of $U=2.0$~eV, $U'=1.3$~eV and 
$U=1.1$~eV, $U'=0.7$~eV. Clearly, in all respects, the agreement gets 
worse with decreasing $U,U'$, strongly supporting the strong 
correlation-based view. This excellent quantitative agreement with 
{\it both} one- and two-particle spectra in the normal state of 
${\rm LaO_{1-x}FeAsF_{x}}$ encourages us to make a deeper analysis of 
transport properties for both La- and ${\rm SmO_{1-x}FeAsF_{x}}$.  

\begin{figure}
%\begin{center}
\includegraphics[width=\columnwidth]{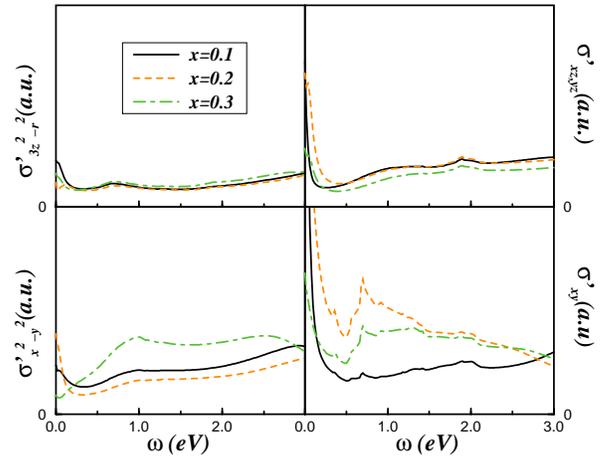}
\includegraphics[width=\columnwidth]{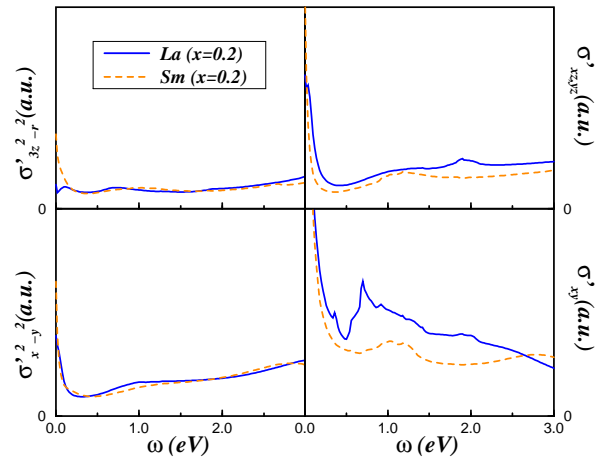}
%\includegraphics[width=3.1in]{F4b-opt.eps}
%\end{center}
\caption{(Color online) Top panel: Orbital-resolved optical conductivity 
of ${\rm LaO_{1-x}FeAsF_{x}}$ for $x=0.1, 0.2, 0.3$, within LDA+DMFT. Apart 
from the very small quasi-coherent component in $\sigma_{xy}(\omega)$, all 
other orbital components exhibit {\it incoherent} non-FL responses, with 
clear wipe-out of the ``Drude'' response at low-energy. Bottom panel: 
Comparison between the orbital resolved optical spectra for doped La- and 
Sm-based Fe-pnictides, showing very similar responses in both cases.}
\label{fig5}
\end{figure}

Using Eq.~(\ref{optc}), we have computed the optical conductivity, which 
is shown in Fig.~\ref{fig5} for all $d$-orbitals for both Fe-pnictides.  
Anisotropic responses, dictated both by LDA band-structure, as well as 
by correlation effects (see below) are clearly visible. A very interesting 
aspect of the results is the observation of strong {\it incoherent} 
features in $\sigma_{a}(\omega)$: a Drude-like peak, with very small 
weight, exists only in $\sigma_{xy}(\omega)$, and is even smaller, 
almost vanishing, in the $xz,yz$ optical response. For all other orbitals, 
the optical response is totally incoherent, with distinct non Drude 
contribution (also see the carrier mass/lifetime results below). Large 
scale SWT across huge energy scales $O(4.0)$~eV with doping is also 
explicit in the results.  In the MO-DMFT, this SWT is driven by the 
{\it dynamical} correlations associated with large on-site interactions 
$U$ and $U'$: the latter causes {\it inter-orbital} SWT on the observed 
scale, and is intimately linked with underlying ``Mottness'' in the 
MO model. The orbital resolved optical conductivity shows a distinctly 
non-Drude component, along with a very slow decrease with increasing 
$\omega$ up to high energy, $O(3.0)$~eV. This is observed clearly in 
Fig.~\ref{fig5} for {\it both} pnictides. At higher energies, one 
expects {\it both}, the inter-orbital transitions, as well as those 
between the bands neglected within the DMFT to start contributing to 
$\sigma_{a}(\omega)$. Thus, one should expect to find good agreement 
with experiment up to $\omega\simeq O(3.0)$~eV.  It is rather satisfying 
to notice that precisely the above features, i.e, non-Drude low-energy 
part, a slowly decaying (in $\omega$) contribution at higher energies, 
and large-scale SWT with doping, have indeed been observed in Fe-pnictides. 
These imply that Fe-pnictides should be considered as strongly correlated 
systems, in the proximity of a MI state.~\cite{si} From the total optical
conductivity, $\sigma(\omega)=\sum_{a}\sigma_{a}(\omega)$, we have estimated 
the average plasma frequency from the sum rule,

\be
\int_{0}^{\infty}\sigma(\omega)d\omega=\frac{\omega_{p}^{2}}{4\pi} \;,
\ee
yielding $\omega_{p}=0.76$~eV, in excellent agreement with the experimental
estimate of $0.68$~eV for ${\rm LaO_{1-x}FeAsF_{x}}$ with $x=0.1$. This is 
a renormalization of a factor of $3$ relative to the LDA estimate.  For 
comparison, we have also computed $\omega_{p}$ with reduced $U,U'$, whence 
$\omega_{p}$ increases smoothly toward its LDA value (this is also visible 
from the reflectivity curves, where the kink moves to higher energy with 
decreasing $U,U'$).    
 
\begin{figure}
%\begin{center}
\includegraphics[width=\columnwidth]{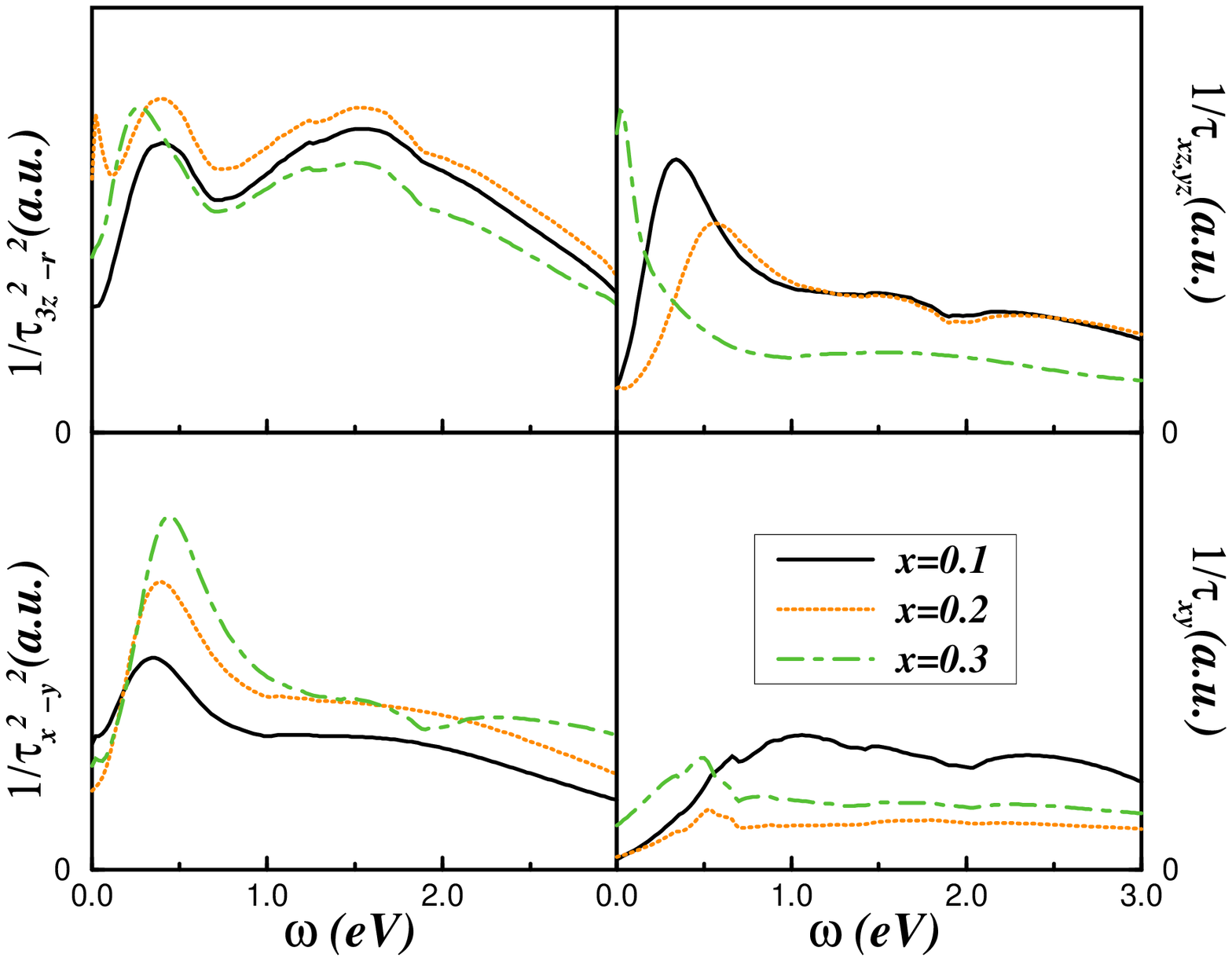}
\includegraphics[width=\columnwidth]{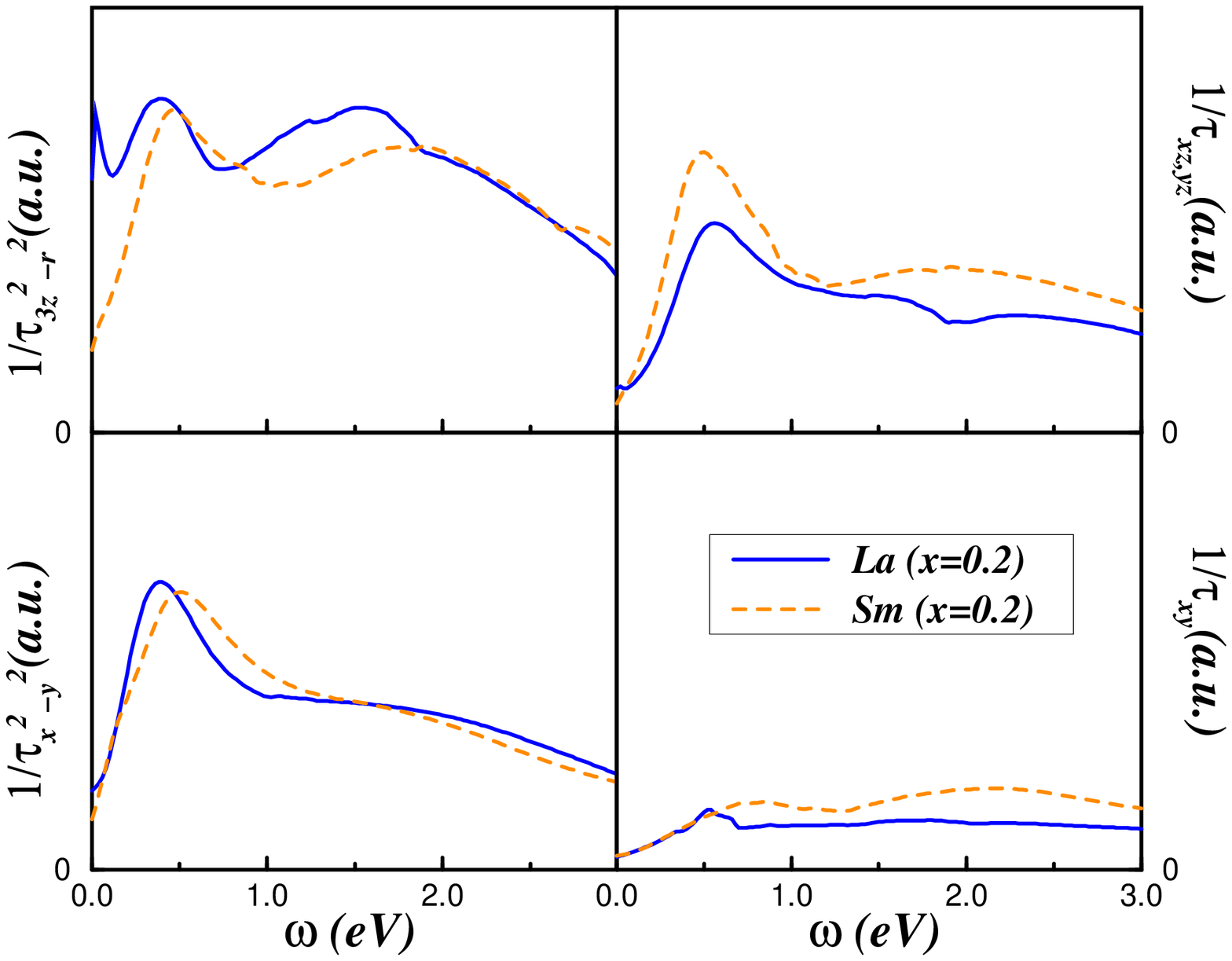}
%\includegraphics[width=3.1in]{F5bt-opt.eps}
%\end{center}
\caption{(Color online) Top panel: Orbital-resolved scattering rates of 
${\rm LaO_{1-x}FeAsF_{x}}$ for $x=0.1, 0.2, 0.3$, within LDA+DMFT. {\it All} 
orbital components exhibit {\it incoherent} non-FL responses, but the 
scattering rate for the $xy$ orbital carriers remains most metallic. For 
the others, the scattering rates exhibit large values at $\omega=0$, in 
full accord with the emergence of low-energy pseudogap-like features in 
their corresponding one-particle and optical lineshapes. Bottom panel: 
Comparison between the orbital resolved scattering rates for doped La- 
and Sm-based Fe-pnictides, showing very similar responses in both cases. 
Notice the enhanced low-energy incoherence for the La-pnictide.}
\label{fig6}
\end{figure}

\begin{figure}
%\begin{center}
\includegraphics[width=\columnwidth]{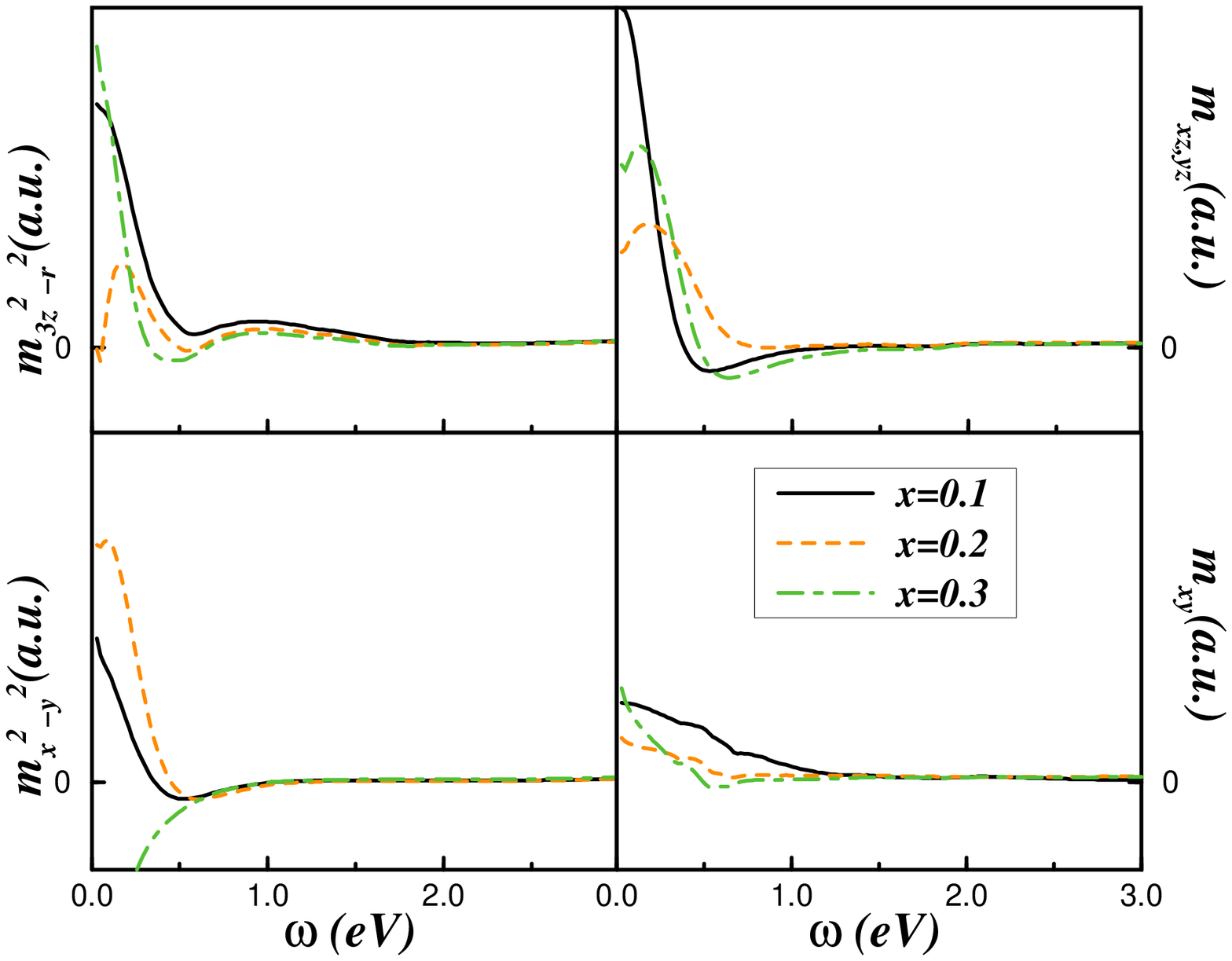}
\includegraphics[width=\columnwidth]{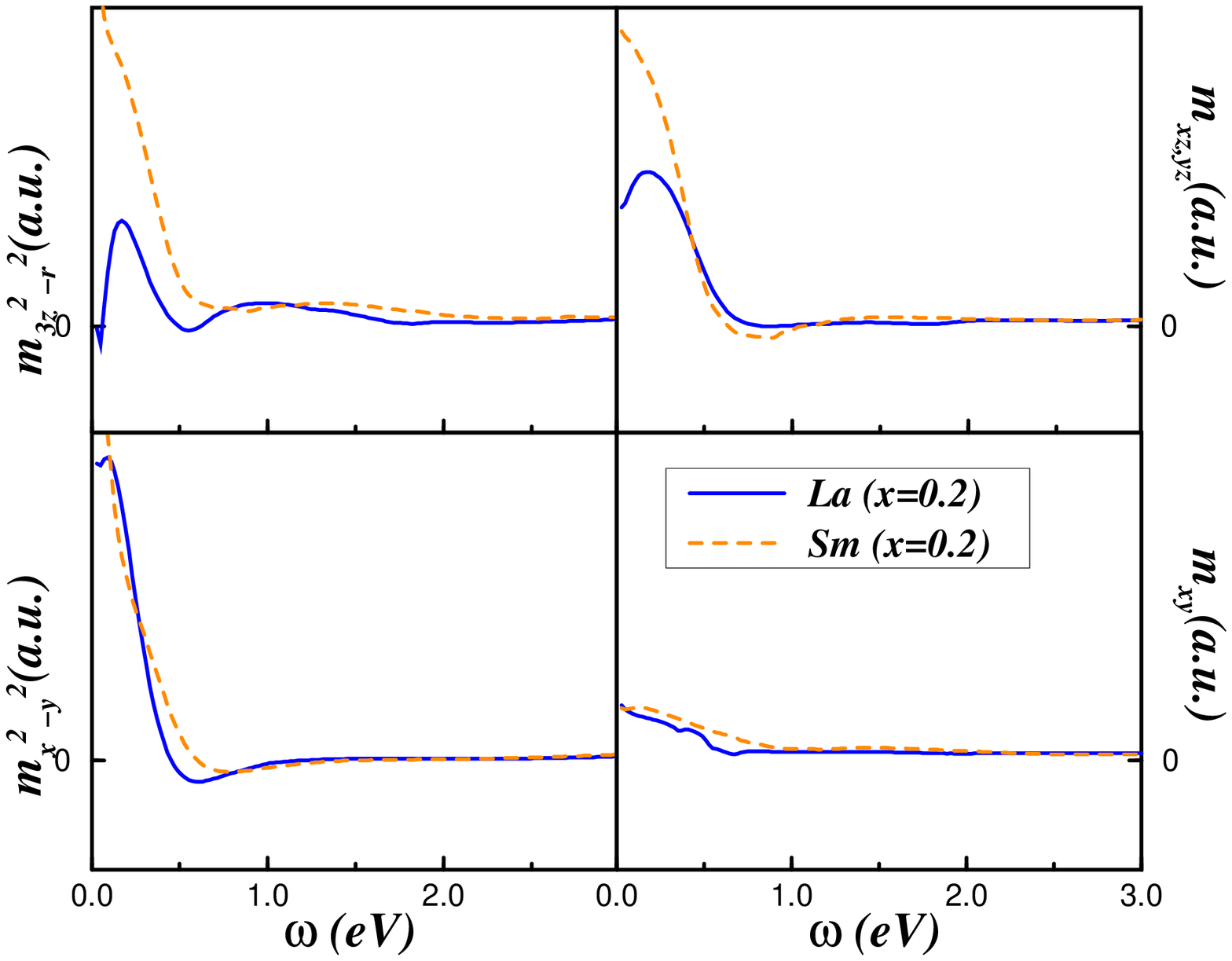}
%\includegraphics[width=3.1in]{F5bm-opt.eps}
%\end{center}
\caption{(Color online)  Top panel: Orbital-resolved dynamical masses of 
${\rm LaO_{1-x}FeAsF_{x}}$ for $x=0.1, 0.2, 0.3$, within LDA+DMFT. {\it All}
orbital components exhibit strong $\omega$-dependence at low energy, in 
line with the incoherent optical responses.  Only the mass for the $xy$
orbital carriers seems to approach a {\it constant}, correlation-enhanced 
value at low energy.  For the others, the dynamical masses exhibit strong 
$\omega$ dependence up to low-energy, in full accord with the emergence of 
low-energy incoherent, or pseudogap-like features in their corresponding 
one-particle and optical lineshapes.
Bottom panel: Comparison between the orbital resolved dynamical masses for 
doped La- and Sm-based Fe-pnictides, showing very similar responses in 
both cases.  Notice the enhanced low-energy incoherence for the La-pnictide.}
\label{fig7}
\end{figure}

Using the extended Drude parametrization allows us to estimate the
frequency-dependent transport scattering rate, $\tau^{-1}(\omega)$, as well
as the dynamical mass, $m^{*}(\omega)$.  In Figs.~\ref{fig6} and~\ref{fig7}
(top panels), we show these quantities for each orbital, $a$, as a function 
of doping.  Distinctive non-FL features are clearly evident. We emphasize 
that these results are valid in the symmetry unbroken metallic state, i.e, 
without SDW or SC order: this is the case above $T_{c}$ in doped 
Fe-pnictides for $x \ge 0.1$. Both $\tau^{-1}$ and $m^{*}$ show distinctly 
{\it orbital-selective} non-FL behaviors. With the exception of the $xy$ 
orbital, $m^{*}(\omega)$ for the other ($x^{2}-y^{2},xz,yz,3z^{2}-r^{2}$) 
orbitals continues to increase in a power-law-like fashion, or in a fashion 
consistent with the onset of {\it incoherent} pseudogap 
behavior,~\cite{[klein]} down to lowest energy. Correspondingly, the 
respective scattering rates clearly exhibit a sublinear $\omega$ dependence 
(with anisotropic, $\omega=0$ values) at low energy. Interestingly, the 
DMFT optical spectra of both La and Sm-based pnictides show that the 
dominant low-energy ``metallic'' contribution comes from the $d_{xz,yz,xy}$ 
bands, while clear {\it pseudogap} response is manifest in the 
$d_{x^{2}-y^{2},3z^{2}-r^{2}}$ channels: the latter are almost Mott 
localized. This observation is consistent with ARPES data,~\cite{[arpes]} 
where the {\it correlated} spectral function  is measured; extant results 
show broadened ``quasiparticle'' bands on $xy,yz,xz$ character crossing 
$E_{F}$. In our DMFT, only the ``more metallic'' $xy,yz,xz$ bands will 
then show up as quasicoherent bands crossing $E_{F}$, while, for the 
pseudogapped $x^{2}-y^{2},3z^{2}-r^{2}$ bands, the extremely large 
scattering rates (with a pseudogap) will obliterate these in ARPES.  

It is instructive to analyze the differences between the optical spectra 
for La- and Sm pnictides.  In the lower panels of Figs.~\ref{fig5},
~\ref{fig6}, and~\ref{fig7}, we explicitly show these. Clearly, the 
anisotropic pseudogap features discussed above are more pronounced for 
the La-based Fe-pnictide. To understand this difference, we recall that 
the LDA band widths for Sm-based Fe-pnictide are about $O(0.5)$~eV wider 
than those for La-based Fe-pnictide, an observation that is consistent 
with the higher chemical pressure induced by the smaller (Sm) ion in 
the former case. Hence, in our DMFT picture, the correspondingly 
narrowed $d$ orbitals for La-based Fe-pnictide will be closer to Mott 
localization vis-a-vis those for the Sm-based Fe-pnictide; this manifests 
itself in the observation of pseudogap signatures in 
$\tau_{a}^{-1}(\omega),m_{a}^{*}(\omega)$ for La pnictide, while these are 
weaker for Sm pnictide. Apart from these material dependent differences, 
the aspect of large normal state {\it incoherence} is clearly reflected 
in the results for {\it both} pnictides.
 
Finally, we remark that, while such extended Drude analysis for the
``single layer'' Fe-pnictides remains to be done with single crystals, a
non-FL scattering rate is indeed extracted from the optical spectra for
doped ${\rm BaFe_{2}As_{2}}$.~\cite{timusk} Based on our results, we
predict that similar incoherence will characterize the normal state optical
response of the single-layer Fe-pnictides as well. The sublinear-in-$\omega$
scattering rate is also consistent with the sublinear-in-$T$ dependence
of the $dc$ resistivity in doped Fe-pnictides~\cite{si} above $T_{c}(x)$.  

\section{Discussion}

The lack of any Drude component in the low-energy optical response implies
that the symmetry-unbroken metallic state above $T_{c}$ in the doped
Fe-pnictides is {\it not} a FL, in the sense that the one-particle
propagators exhibit a branch cut structure, rather than a renormalized
pole structure, at low energies.

What is the microscopic origin of the non-FL features found in the DMFT
solution?  In MO systems, the orbital-resolved hopping matrix elements
(diagonalized in the LDA) are very directional, being sensitive functions
of orbital {\it orientation} in the real structure.  Further, the $d$ bands
are shifted relative to each other because of the action of the crystal
field (of $S_{4}$ symmetry in Fe-pnictides), and the six $d$ electrons are
distributed among all $d$ orbitals.  In this situation, strong, MO
correlations cause two, intimately linked changes:

$(i)$ the static MO-{\it Hartree} shift, which depends upon the occupations 
of each orbital, as well as on the inter-orbital $U'$ and $J_{H}$, 
renormalizes the on-site energies of each orbital in widely different ways. 
In particular, it causes inter-orbital charge transfer between the various 
$d$-orbitals, self-consistently, modifying their energies and occupations. 
This effect is also captured in LDA+U approaches. Specifically, the 
lower-lying orbital(s) in LDA are pushed further down by the MO-Hartree 
shift, the amount of which is determined by their occupation(s) and by 
the values of $U',J_{H}$ relative to their respective LDA band width(s), 
and to the {\it bare} crystal field splitting in LDA.

$(ii)$ More importantly, the dynamical correlations associated with 
$U,~U',~J_{H}$ results in a large-scale transfer of dynamical spectral 
weight.  Small changes in the LDA band structure induced by $(i)$ above 
(or by changes in external perturbations in general) induce large 
{\it changes} in SWT, drastically modifying LDA lineshapes.

$(iii$) Crucially, the renormalized Fermi energy is computed selfconsistently
in DMFT by requiring consistency with the Freidel-Luttinger theorem; i.e,
$E_{F}$ is computed by demanding that the renormalized Fermi surface
encloses the total number of $d$-electrons in the system, as long as no 
broken symmetry states are considered.

Generically, as $U'$ increases ($J_{H}$ is usually fixed for a given
$d$-state), the lower-lying subset of $d$-orbitals gets selectively Mott
localized; the metallic phase is then the OS metal found recently in various 
contexts.~\cite{[mllc],[georges]} Once this selective localization occurs 
within the DMFT, the low energy physical response is governed by strong 
scattering between the effectively Mott-localized and the renormalized, 
itinerant components of the matrix spectral function.  The problem is thus 
effectively mapped onto a Falicov-Kimball type of model, as has been noticed 
in earlier work.~\cite{[georges],[mllc]} Within DMFT, the itinerant fermion 
spectral function then shows a low-energy pseudogapped form, while the 
``localized'' spectral function shows a power-law fall-off as a function 
of energy, as long as the renormalized $E_{F}$ is {\it pinned} to the 
renormalized orbital energy of the localized orbital(s). This is understood 
from the mapping of the corresponding impurity model to that of the ``X-ray
edge'',~\cite{[pwa-1967]} where the orthogonality catastrophe destroys FL
behavior.~\cite{[nozieres]} The spectral functions then exhibit asymmetric
continuua (branch cuts) at low energy, instead of symmetric Abrikosov-Suhl
Kondo resonance features, and the metallic phase is {\it not} a FL.  This
incoherence is mirrored in the optical responses in $D=\infty$, since the
optics is entirely determined by the {\it full} DMFT one-particle Green
functions in this limit.  This is entirely consistent with our results,
and strongly suggests that effects akin to the orthogonality catastrophe,
arising from strong, incoherent, inter-orbital scattering, produce the
incoherent metal behavior in the ``normal'' state of Fe-pnictides.      

This loss of one-particle coherence corresponds to drastic reduction of
the carrier kinetic energy, and implies the {\it irrelevance} of
one-particle terms in the RG sense.~\cite{[kot-rg]} Interestingly, this 
clears the way for two-particle instabilities to take over as $T$ is 
lowered; they pre-empt the non-FL behavior from persisting down to $T=0$. 
This could be either:

$(i)$ the ${\bf q}=(\pi,0)$ SDW for $x<x_{c}$. This is in fact further
reinforced by the near-nested character of the electron- and hole Fermi
pockets in the Fe-pnictides, already visible in weak coupling RPA
calculations.~\cite{raghu,[chubukov]} The SDW instability  at $x=0$ is 
then interpretable as an instability in the particle-hole channel, aided 
by near FS nesting. That AF-SDW survives the lack of such nesting features 
away from $x=0$ is an observation which favors a strong coupling scenario. 
An intermediate-to-strong coupling version of a similar instability to 
exactly the same state results from the LDA+DMFT work of 
Haule {\it et al.}~\cite{Haule} A strong coupling method as used here 
will yield the same AF instability, in analogy with one-band Hubbard model 
studies, where the AF ordered state at half-filling remains the Neel ordered 
state, evolving from the Slater type to the Mott-Hubbard type as $U/t$ is 
increased. Though weak-coupling RPA approaches, valid for small $U/t$, can 
indeed give the correct magnetic {\it ground} states for $U\ll W(=2zt)$, 
the normal state incoherence characteristic of Fe-pnictides, which are now 
identified to be in the intermediate coupling limit ($U\simeq W$), is beyond 
their scope. This is because the incoherent part of $G_{a}({\bf k},\omega)$ 
is completely neglected there, and one deals with propagators having only a 
coherent QP pole structure at low energy. This is valid for small $U$, but 
not for $U>W$, the non-interacting bandwidth.

$(ii)$ An unconventional SC for $x>x_{c}$. This corresponds to an instability 
in the particle-particle channel. Independent of the precise order parameter 
symmetry, a matter of intense debate,~\cite{si,[3],ere,kuroki} the 
observation of small superfluid density, short-coherence length, large 
$2\Delta/kT_{c}$ ratios, along with large energy scale changes in optical 
response across $T_{c}$, are all hallmarks of a SC closer to the Bose 
condensation limit, rather than the weak coupling BCS limit.
    
\begin{figure}
%\begin{center}
\includegraphics[width=\columnwidth]{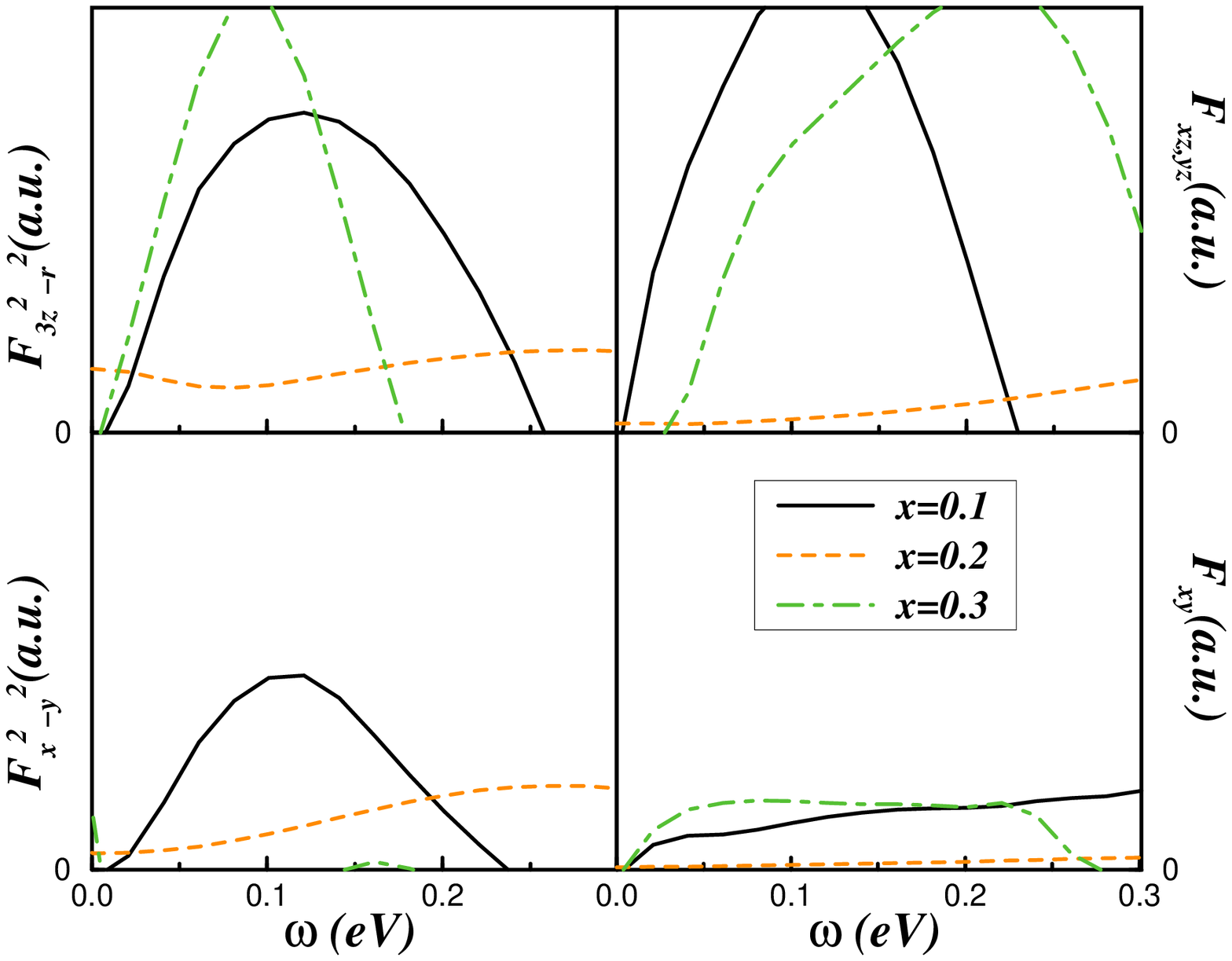}
\includegraphics[width=\columnwidth]{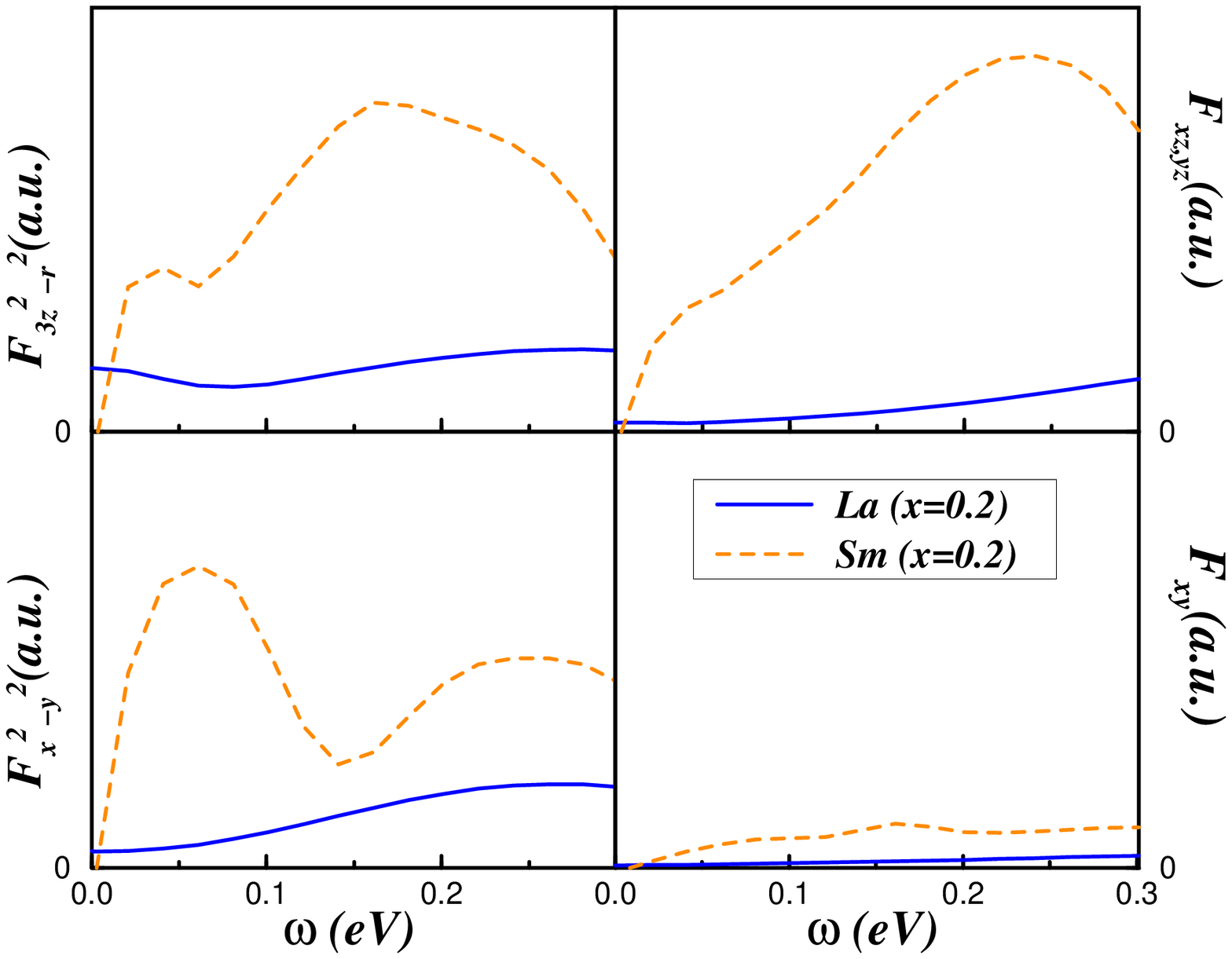}
%\includegraphics[width=3.1in]{F6b-opt.eps}
%\end{center}
\caption{(Color online)  Top panel: Orbital-resolved ``glue'' functions for 
${\rm LaO_{1-x}FeAsF_{x}}$ for $x=0.1, 0.2, 0.3$, within LDA+DMFT. {\it All}
orbital components contribute, albeit anisotropically, at low energy. This 
suggests that the instability to superconductivity in Fe-pnictides will 
involve multiple bands. The lack of any sharp feature shows that the ``glue''
is an overdamped electronic continuum (see text).  The two-peak structure 
at low energy is reminiscent of what is seen in high-$T_{c}$ cuprates, and
has a multi-orbital character. Bottom panel: Comparison between the ``glue''
functions for doped La- and Sm-based Fe-pnictides, showing similar incoherent 
electronic continuum features in both cases. }
\label{fig8}
\end{figure}

Our findings are in accord with  Si {\it et al.}~\cite{si} and Haule
{\it et al.},~\cite{Haule} who, along with Baskaran,~\cite{bask} were
the first to recognize the strong coupling aspect of Fe-pnictides. On the
other hand, several works address both AF and SC within weak coupling
scenarios.~\cite{[chubukov],raghu} In the latter, the ``normal'' state
above $T_{c}$ is a FL metal, and SC arises via a BCS like instability of
this MO (multi-band) FL.  In the strong correlation limit, however, the 
``normal'' state is itself incoherent, and so there are no FL-like QPs 
to pair up into usual BCS-like cooper pairs. In other words, SC must 
then arise as an instability of an unusual metallic state without FL QPs, 
or, to rephrase it, directly from overdamped, collective multi-particle 
modes. The {\it frequency}-dependence of the self-energy is important in 
the latter case, and consideration of the instability of such an incoherent 
state to the SC state leads to a strongly frequency-dependent SC gap 
function. In such a ``strong coupling'' picture, the physical response 
functions {\it in} the SC state are controlled by both, the symmetry of 
the SC order parameter, as well as the strong, $\omega,T$ dependent damping 
originating in the incoherent ``normal'' state. Observation of features 
such as the absence of the Hebel-Slichter peak in the NMR 
$T_{1}^{-1}(T)$~\cite{[3]} and the large scale modification of the optical 
spectral weight across $T_{c}$~\cite{vdm} in the Fe-pnictides are strongly 
suggestive of such a strong coupling scenario. These features are again 
reminiscent of cuprates,~\cite{rev_hTc,vandm} where the role of ``normal'' 
state (non-FL) incoherence is well documented. To make their role in the 
Fe-pnictides more explicit, we use our optical results to show the pairing 
``glue'' function for Fe-pnictides in Fig.~\ref{fig8}. In line with the 
conclusions extracted from the analysis of the incoherent optical 
conductivity above, ${\cal F}(\Omega)$ shows interesting features. We find

$(i)$ a two ``peak'' structure, with both peaks strongly broadened by
incoherent scattering. We emphasize that this broad, two-peak structure
arises from the incoherent one-particle propagators, and represents a
multiparticle electronic continuum.  This could be interpreted as
overdamped ``bosonic'' modes, if one associates the short ranged, strongly
coupled spin-orbital modes with the incoherent, short-distance components
of the conventional bosonic modes used in the itinerant descriptions. It
is interesting to notice that similar features, namely, strong non-FL
signatures in optics, as well as a strongly damped, two-peaked, low-energy
continuum is also characteristic of high-$T_{c}$ cuprates up to optimal
doping.~\cite{rev_hTc,vandm} Thus, our results show that SC arises from 
an incoherent normal state, and that coupling carriers to an incoherent 
electronic continuum pairs them up in the SC. Hence, we propose that the 
underlying Mott-Hubbard physics known to underpin the anomalous responses 
in cuprates is also at work in the Fe-pnictides.
    
$(ii)$ {\it All} $d$-orbital components show up in ${\cal F}_{a}(\omega)$,
albeit anisotropically. This supports a MO (multi-band) origin for SC
pairing in Fe-pnictides. Several interesting features are visible upon
close scrutiny: the ``glue'' function is larger for those (orbitals) bands
which are closer to Mott localization, as seen by comparing the respective
curves in ${\cal F}_{a}(\Omega)$ and $\rho_{a}(\omega)$, while the most
``itinerant'' $xy$ band has the smallest contribution to ${\cal F}$.
In other words, there is pronounced orbital-induced anisotropy in the
pairing ``glue'' function. This should imply an orbital-dependent, multiple
gap SC, which may not be inconsistent with recenttheoretical 
indications.~\cite{[5],[chubukov]} In other words, suppression of 
one-particle coherence (in the DMFT propagators, $G_{a}({\bf k},\omega)$)
makes two-particle (collective) processes more relevant, by enhancing the
respective collective susceptibilities in the charge and spin channels.  
In the doped case, lack of nesting features in the electron-like and
hole-like Fermi sheets suppresses the ${\bf q}=(\pi,0)$ SDW found for
$x=0$, leaving multi-band SC as the only relevant two-particle instability.
In such a MO-SC, opening up of an orbital-dependent gap should be
linked to orbital-dependent coupling of the carriers to the electronic 
``glue'' via ${\cal F}_{a}(\omega)$: remarkably, a very recent 
indeed shows exactly this feature.~\cite{[Richard]} This agrees 
qualitatively with our expectations from a strongly orbital-dependent 
``glue'' function, as derived above. It is also interesting to note that 
a recent ARPES study on ${\rm (Sr/Ba)_{1-x}K_{x}Fe_{2}As_{2}}$~\cite{[wray]} 
also finds quasiparticle kinks in the quasiparticle {\it dispersion} in
the binding energy range from $15$~meV to $50$~meV, again reminiscent of what
is observed in high-$T_{c}$ cuprates,~\cite{[fink]} but also in {\it three}
dimensional, correlated systems like ${\rm SrVO_{3}}$.~\cite{[vollha]}  
The latter fact points to its connection to underlying Mottness 
characteristic of a correlated system. In our DMFT, we recall that the 
orbital-resolved DOS show low-energy kinks precisely in the $15-50$~meV 
range (see Fig.~\ref{fig2}). These are attributed to low-energy, collective, 
inter-orbital fluctuations,~\cite{SmFe} and we have shown that the 
resulting LDA+DMFT DOS gives excellent quantitative agreement with the 
normal state kink-like feature in angle {\it integrated} PES.~\cite{[pesa]} 
Based on our calculation, we propose that an ARPES measurement done for
${\rm (La/Sm)O_{1-x}FeAsF_{x}}$ should uncover kink structure(s) in a similar 
low-energy ($20$~meV) range.  
 
Whether additional density-wave instabilities may also interfere with SC is
an interesting issue.  Two-band model studies~\cite{[kivelson],[chubukov]} 
do suggest additional channels involving nematic or current instabilities:
whether they can seriously compete with SC or remain sub-leading in the 
full five-band model, is still unclear.  In any case, our analysis shows 
that consideration of {\it all} $d$-orbitals is necessary for a proper
microscopic description of multi-band SC in Fe-pnictides. We leave the 
detailed consideration of the instability of the MO, incoherent 
metal found here to a MO SC for future consideration.

A brief discussion of the similarities and differences between cuprates and
Fe-pnictides is in order at this point.  The predominance of normal state
incoherence in cuprates is well-documented by now,~\cite{[pwa-book]} and its
link with the strong on-site electronic correlations is well known.  In
particular, strong non-FL features (with a branch cut form of the
one-particle propagator, $G({\bf k},\omega)$) in ARPES, $dc$ and $ac$ 
transport, magnetotransport, as well as in magnetic fluctuations bears 
this out in a very remarkable way.  Up to optimal doping, the cuprates 
are not describable in terms of the Landau FL picture, and the role of 
strong Mottness and short-ranged AF spin correlations in this context is 
recognized.  In the underdoped cuprates the instability to the $d$-wave 
SC state is quite far from the weak-coupling BCS variety: large-scale 
redistribution of spectral weight across $T_{c}$ as well as strong vortex 
liquid fluctuations, short SC coherence lengths, and small superfluid 
stiffness, among other observations, put this transition closer to the 
Bose condensed limit.

Many of these features, like normal state incoherence, along with pseudogap
behavior in {\it both} charge and spin fluctuations, are also characteristic
of Fe-pnictides.  Further, in single-layer Fe-pnictides, the superfluid 
density is small, the upper-critical fields, $H_{c2}$, is high, the SC
in-plane coherence length is short ($\xi_{pair}<20$~{\AA})~\cite{[wray]} and 
appreciable redistribution of spectral weight across $T_{c}$ is visible. 
These similarities then strongly support the strong correlation-based view 
for Fe-pnictides as well.

There are noticeable differences between the cuprates and the Fe-pnictides.
First, SC in cuprates most likely involves a single, strongly $p-d$
hybridized band with strong electronic correlations, and arises upon 
doping a Mott-Hubbard insulator with AF order.  In contrast, SC in 
Fe-pnictides is most probably of the multi-band variety, and arises 
upon doping a very bad metal, which may be close to a Mott-Hubbard 
insulating state.~\cite{si} This means that the microscopic electronic 
``glue'' for pairing in both cases will be quite different. Nevertheless, 
in view of the underlying relevance of Mottness in both cases, one may 
expect more similarities in the {\it structure} of the SC instability 
in both cases. It is more likely, given the explicit MO situation in 
Fe-pnictides, that additional competing instabilities may be at work. 
In particular, the nematic and current instabilities, which have been 
invoked as competitors of $d$-wave SC in cuprates,~\cite{metva} might
also play a role here.~\cite{[kivelson],[chubukov]} As remarked early by
Baskaran,~\cite{bask} it is possible that being able to think about
situations where these competing instabilities could be suppressed can
push up the SC $T_{c}$ in Fe-pnictides to even higher values.

Our analysis here has been carried out for the symmetry-unbroken metallic
state of the Fe-pnictides.  At low $T$, this incoherent state becomes 
unstable to either AF-SDW order with ${\bf Q}_{SDW}={\bf q}=(\pi,0)$ or 
to SC order, depending on $x$, though some studies also suggest co-existence 
of the two orders. As discussed in literature,~\cite{si,[yild]} the AF-SDW 
state involves strong geometric frustration (GF) in the inter-orbital 
hopping matrix elements. Characterization of magnetic fluctuations has 
indeed been carried out, both in the strong coupling ($J_{1}-J_{2}$ 
Heisenberg model),~\cite{[sachdev]} as well as within weak coupling HF-RPA 
work.~\cite{ere} Observation of features akin to high-$T_{c}$ cuprates 
found experimentally, as discussed above, put the Fe-pnictides into the 
strongly correlated category, though not so strongly correlated as cuprates, 
which are doped Mott insulators. While the fact that we are able to describe 
{\it both} the one-electron responses and the reflectivity of La-pnictide 
strongly suggests that LDA+(MO)DMFT is adequate for a quantitative 
description of these correlation effects, we expect geometrical frustration 
effects to become relevant at low $T<T_{N}(x)$, at least for the SDW phase. 
In view of the fact that the spatial extent of correlations is short in GF 
systems, we remark that the dynamical effects associated with these 
short-ranged magnetic correlations may not change the above results 
substantially. They will certainly not modify them qualitatively; indeed, 
one would expect that additional consideration of the dynamical effects 
of such short-ranged, strongly frustrated couplings 
($J_{1},J_{2}$ with $J_{2}/J_{1}\simeq O(1)$)~\cite{[yild]} would push 
the almost totally incoherent normal state derived above even more toward 
incoherence. 
 
\section{Conclusion}

In conclusion, we have studied the normal state electrodynamic response of
two Fe-pnictides, ${\rm La_{1-x}FeAsF_{x}}$ and ${\rm SmO_{1-x}FeAsF_{x}}$,
within the LDA+DMFT formalism.  Armed with the very good agreement
with one-electron responses (PES and XAS) between experimental and LDA+DMFT
spectral functions, we {\it also} find very good quantitative agreement
with published reflectivity data on ${\rm LaO_{1-x}FeAsF_{x}}$.  Building
upon these agreements, we have studied the optical response in detail. We 
find an incoherent optical response, with strongly $\omega$-dependent 
effective masses and transport lifetimes (scattering rates) at low energy. 
These features are linked to the incoherent, near pseudogap-like features 
found in the orbital resolved, one-particle spectral functions in LDA+DMFT. 
The very good agreement found between theory and experiment for {\it both} 
pnictides is strong evidence for the relevance of ``Mottness'', i.e, to 
proximity of the normal state of Fe-pnictides to a correlation-driven 
Mott-Hubbard insulating state. Further, based on an estimation of the 
``glue'' function, we propose that this should be understood as arising 
from a multi-particle electronic continuum, which could also be interpreted 
in terms of an overdamped ``bosonic'' glue. Finally, {\it all} $d$-orbitals 
should contribute to SC pairing, albeit anisotropically, and one should 
have an orbital-dependent, multiple-gap SC. It is an interesting task to 
investigate the low-$T$ instabilities of this incoherent metallic state: 
we leave this for the future.

%\acknowledgements
L.C. thanks S.-L. Drechsler for discussions. 
%and IFW Dresden for financial support. 
S.L. acknowledges ZIH Dresden for computational time. M.S.L. thanks 
the MPIPKS, Dresden for financial support.

\end{document}